\documentclass[12pt,preprint]{aastex}
 
\def\bi{\begin{itemize}}
\def\ei{\end{itemize}}
\def\be{\begin{equation}}
\def\ee{\end{equation}}

\def\gtrsim{\mathrel{\hbox{\rlap{\hbox{\lower4pt\hbox{$\sim$}}}\hbox{$>$}}}}
\def\lesssim{\mathrel{\hbox{\rlap{\hbox{\lower4pt\hbox{$\sim$}}}\hbox{$<$}}}}
\def\gtrsim{\mathrel{\hbox{\rlap{\hbox{\lower4pt\hbox{$\sim$}}}\hbox{$>$}}}}
\def\lesssim{\mathrel{\hbox{\rlap{\hbox{\lower4pt\hbox{$\sim$}}}\hbox{$<$}}}}
\slugcomment{Draft \today}

\shortauthors{Cooper, Lister and Kochanczyk}
\shorttitle{MOJAVE - I. VLA 1.4 GHz Images}
\begin{document}

\title{MOJAVE: Monitoring of Jets in AGN with VLBA Experiments. III. Deep VLA 
Images at 1.4 GHz }

\author{N. J. Cooper}
\affil{Department of Physics, Purdue University, 525 Northwestern
Avenue, West Lafayette, IN 47907}
\email{ncooper@physics.purdue.edu}

\author{M. L. Lister}
\affil{Department of Physics, Purdue University, 525 Northwestern
Avenue, West Lafayette, IN 47907}
\email{mlister@physics.purdue.edu}
 
\author{M. D. Kochanczyk}
\affil{Department of Physics, University of Texas at Austin,
 1 University Station C1600, Austin, TX 78712}
\email{martin@physics.utexas.edu}

\begin{abstract}

The MOJAVE blazar sample consists of the 133 brightest, most compact
radio-loud AGN in the northern sky, and is selected on the basis of
VLBA 2 cm correlated flux density exceeding 1.5 Jy (2 Jy for
declinations south of 0$^\circ$) at any epoch between 1994 and 2003.
Since 1994 we have been gathering VLBA data on the sample to measure
superluminal jet speeds and to better understand the parsec$-$scale
kinematics of AGN jets.  We have obtained 1.4 GHz VLA$-$A
configuration data on 57 of these sources to investigate whether the
extended luminosity of blazars is correlated with parsec$-$scale jet
speed, and also to determine what other parsec$-$scale properties are
related to extended morphology, such as optical emission line strength
and gamma$-$ray emission.  We present images and measurements of the
kilo-parsec scale emission from the VLA data, which will be used in
subsequent statistical studies of the MOJAVE sample.

\end{abstract}
 
\ 
\ 
\section{Introduction}

Blazars are an extreme class of radio source that represent only a small 
percentage of active galactic nuclei (AGN), yet dominate the sky at high 
energies.  They are characterized by highly relativistic jets that are 
oriented toward us with a very small angle of incidence.  This often 
results in apparent superluminal jet motions that are caused by a Doppler  
compression in the apparent time duration of signals from the source.  This 
creates an ideal set of circumstances for studying the kinematics of jets, 
since many decades of evolution are often compressed into much shorter timescales.  
MOJAVE (Monitoring of Jets in Active galactic nuclei with VLBA Experiments) 
is a long term program that seeks to better understand the physics of AGN on the 
parsec$-$scale via regularly$-$spaced, high resolution VLBA observations 
(\citealt{K98, KL04, LM05}).  The MOJAVE sample contains 133 sources satisfying 
the criteria, 
\bi

\item J2000 Declination $>$ $-$20$^\circ$
\item Galactic latitude $|b| > 2.5^\circ$
\item VLBA 2 cm correlated flux density exceeding 1.5 Jy (2 Jy for declination
south of 0$^\circ$) at any epoch between 1994 and 2003. 

\ei
By selecting on the basis of compact flux density at a short wavelength, the
sample favors highly beamed relativistic jets (i.e., blazars).  This makes it
well suited for quantifying the strong selection effects associated with blazar
samples \citep{LM97}.

An important goal of the MOJAVE program is to gain a better understanding of 
how AGN jets evolve from parsec to kiloparsec$-$scales. Parsec$-$scale jet 
speed information on the sample is nearly complete (Lister et al., in 
preparation), and a distinct upper envelope is seen in a plot of maximum 
observed jet speed versus parsec$-$scale jet luminosity \citep{CE07}.  
The shape of this envelope is suggestive of an intrinsic correlation between 
jet speed and synchrotron luminosity.  Early hydrodynamic models 
(\citealt{BR74}) predict such behavior, and if verified, would prove useful for
constraining more recent numerical models of jet production that include full 
magneto-hydrodynamic (MHD) and general relativistic (GR) effects 
(e.g., \citealt{KMS04}).
 
The interpretation of the envelope is complicated by the fact that the apparent
luminosity of blazar jets is strongly affected by the bulk Doppler factor 
($\delta$), of the flow.  The latter quantity is dependent on the Lorentz factor 
and viewing angle of the jet, which also affect the apparent speed.  
An alternate means of testing the intrinsic speed/luminosity correlation is to 
compare the apparent jet speeds with extended (radio lobe) emission on 
kiloparsec$-$scales.  The extended emission is not appreciably beamed and is 
well-correlated with the total jet power \citep{GEA88}.  

Several extended flux density surveys that include blazars have been carried 
out with the VLA (e.g., \citealt{AU85, MBP93, C99, RS01}). However, none of 
these samples were as large or complete as MOJAVE, and they lacked 
parsec$-$scale kinematic information.  In order to measure the relatively 
weak extended radio emission in the presence of a strong, highly beamed  blazar 
core component, images with good angular resolution and very high dynamic range 
are necessary.  Although better resolution can be achieved at higher observing 
frequencies, the extended emission becomes faint because of the steep spectral 
index.  We have therefore sought to obtain deep VLA$-$A configuration images of 
the entire MOJAVE sample at 1.4 GHz, which represents a reasonable compromise 
between angular resolution and sensitivity to extended emission.  Here we 
present images of 57 MOJAVE sources; data on the remaining objects (from the VLA 
archive and other observations) will be presented in an upcoming publication.

\section{Observations}

The observations were made with the VLA in the A configuration, which
yields a typical restoring beam of 1.5\arcsec (FWHM), at an observing
frequency of 1.4 GHz.  To minimize the possible effects of radio
interference, and to improve the dynamic range, the data were recorded
in spectral line mode, with 8 channels per IF.  Right$-$ and
left$-$hand circular polarization data at 1.36 GHz and 1.44 GHz were
each assigned to one of four IFs, each having bandwidth of 25 MHz. The
observations took place during filler time on six separate dates
between 2004 September 19 and 2004 November 24 (Table 1). To maximize
{\it(u,v)} coverage, two scans of approximately five$-$minutes
duration, separated in hour angle, were made on each source.  Two
sources, 0059$+$581 and 0109$+$224, had less than 10 minutes
integration time, due to scheduling constraints.  Data on three high
declination sources (0016$+$731, 1150$+$812, 1459$+$718) were not
analyzed due to poor {\it(u,v)} coverage.

Calibration of the data was performed in the AIPS \footnote{AIPS is copyrighted by
Associated Universities, Inc. using the GNU copyright form.} and Difmap
\footnote{DIFMAP was written by Martin Shepard at Caltech, and is part of the
Caltech VLBI software package.} packages. Initial phase calibration
was carried out in AIPS using the most compact sources observed during
each run, with further iterations of imaging and self-calibration done
on each source in DIFMAP. Careful attention was paid to image over the
entire extent of the VLA primary beam, in order to mitigate the
effects of fainter sources in the field on the dynamic range. This
procedure yielded typical r.m.s. noise levels of 0.14 $\mathrm{mJy \;
beam^{-1}}$ and dynamic ranges of $\sim 15 000:1$ (Figure 1).  For each source,
the flux density of the core component 
($S_{core}$) was determined in AIPS by fitting a Gaussian component to the 
image with JMFIT. The size of the fitted Gaussians (FWHM) were in close 
agreement with the size of the restoring beam.  The total flux density was 
determined by summing the flux density within a box fitted around the extent 
of the source.  The extended flux density of the source was calculated using
$S_{ext}=S_{total}-S_{core}$.  The r.m.s. noise was measured in a
blank sky region well away from the source. These values are listed in
Table 2, and the distribution of extended flux density values is shown
in Figure 2.

The images of several sources (0202+149, 0212+735, 0552+398, 0642+449,
1324+224, 1417+385 and 2209+238) required futher analysis to confirm
the presence of very faint extended emission.  The core component was
subtracted from the data, and the residual image was inspected for
emission above 3 times the r.m.s. noise level.  Extended emission was
found for all but two sources, 0642+449 and 2209+238.  For these two
sources we set an upper limit on extended emission equal to three
times the r.m.s. noise.  In these and all the other sources, we cannot
rule out the presence of low$-$level diffuse emission on larger
angular scales than the restoring beam, however.

\section{Discussion}

The sources in the MOJAVE sample appear highly core$-$dominated, since
they are selected on the basis of compact VLBA emission, and not total
(single$-$dish) flux.  Indeed, nearly half of the objects we have
imaged have less than 50 mJy in extended emission at 1.4 GHz, and only
one source ($1222+216$) has $S_{ext} > S_{core}$.  In Figure 3 we show
a histogram of the logarithm of core$-$to$-$extended luminosity ratio
(R$_{c}$).  These values were K$-$corrected to the rest frame of each
AGN using redshifts from the NASA Extragalactic Database assuming
spectral indices of $\alpha = 0$ and $\alpha = -0.7 \quad (S \propto
\nu^\alpha)$ for the core and extended emission respectively.  For
sources without redshift information, z = 1 was assumed.  The shape of
the distribution is similar to that of another core$-$dominated AGN
sample studied by Murphy et al. (1993), with a broad peak centered at
$R_c \simeq 10$.

\section{Conclusions}

We have obtained high$-$dynamic range VLA$-$A configuration images at 1.4 GHz 
of 57 AGN in the MOJAVE sample as part of a project to investigate a 
possible correlation between extended jet luminosity of blazars and 
parsec$-$scale jet speed.  We found extended emission above the $\sim$0.3 mJy
level in all but two sources.  A more detailed analysis of the full MOJAVE 
sample, using arcsecond$-$scale radio images from previous studies, the VLA 
archive and other observations, will be presented in an upcoming paper.

\acknowledgements
 The authors wish to acknowledge the contributions of the other
 members of the MOJAVE project team: Hugh and Margo Aller, Tigran
 Arshakian, Steve Bloom, Marshall Cohen, Dan Homan, Matthias Kadler, 
 Ken Kellermann, Yuri Kovalev, Andrei Lobanov, Eduardo Ros, Rene Vermeulen, 
 and Tony Zensus.

 This research was supported by NSF grant 0406923-AST and the Purdue
 Research Foundation, and made use of the following resources: The
 NASA/IPAC Extragalactic Database (NED), which is operated by the Jet
 Propulsion Laboratory, California Institute of Technology, under
 contract with the National Aeronautics and Space Administration.  The
 Very Large Array (VLA), which is operated by The National Radio
 Astronomy Observatory (NRAO).  The NRAO is a facility of the National
 Science foundation, operated under cooperative agreement with
 Associated Universities, INC.

\clearpage
 



\begin{deluxetable}{cccc} 
\tablecolumns{4} 
\tabletypesize{\scriptsize} 
\tablewidth{0pt}  
\tablecaption{\label{coretable}Log of VLA Observations}  
\tablehead{\colhead{Observation } & \colhead{Number of Sources } 
& \colhead{Flux Density} & \colhead{Number of } \\

\colhead{Date} &\colhead{Observed} & \colhead{Calibrator} & \colhead{VLA Antennas} \\

\colhead{(1)} & \colhead{(2)} & \colhead{(3)} & \colhead{(4)}}
 
\startdata 

2004 Sep 19 & 21 & 0134+329 (3C 048) & 24 \\
2004 Nov 03 & 7  & 0134+329 (3C 048) & 25 \\
2004 Nov 09 & 9  & 0134+329 (3C 048) & 25 \\
2004 Nov 20 & 6  & 1328+307 (3C 286) & 25 \\
2004 Nov 21 & 12 & 0134+329 (3C 048) & 24 \\
2004 Nov 24 & 5  & 0538+498 (3C 147) & 24 \\

\enddata 

\end{deluxetable} 





\begin{deluxetable}{ccccccccr} 
\tablecolumns{9} 
\tabletypesize{\scriptsize} 
\tablewidth{0pt}  
\tablecaption{\label{coretable}Image and Source Properties at 1.4 GHz }  
\tablehead{\colhead{Source}& \colhead{z} &\colhead{S$_{core}$} &\colhead{S$_{ext}$} & \colhead{R.M.S.} &  
\colhead{Lowest Contour} & \colhead{B$_\mathrm{maj}$} & \colhead {B$_\mathrm{min}$} & \colhead{B$_\mathrm{PA}$}  \\ 

\colhead{} & \colhead{} &\colhead{ (Jy)} & \colhead{(Jy)} &   \colhead{(mJy $\rm{beam}^{-1}$)} & \colhead{(mJy $\rm{beam}^{-1}$ )}& \colhead{(arcsec)} & \colhead{(arcsec)}   &  \colhead{ (degrees)}   \\

\colhead{(1)} & \colhead{(2)} & \colhead{(3)} & \colhead{(4)} &  
\colhead{(5)} & \colhead{(6)} & \colhead{(7)} & \colhead{(8)} & \colhead{(9)}} 
\startdata 
0003$-$066 & 0.347 & 2.637 & 0.065 & 0.19 & 1.04 & 1.92 & 1.41 & 23.6 \\
0007+106 & 0.0893 & 0.076 & 0.017 & 0.04 & 0.38 & 1.62 & 1.46 & 43.6 \\
0059+581 & 0.643 & 1.567 & 0.028 & 0.16 & 0.78 & 1.64 & 1.35 & $-$45.1 \\
0106+013 & 2.107 & 2.787 & 0.552 & 0.16 & 0.69 & 1.64 & 1.49 & 9.6 \\
0109+224 & $-$  & 0.363 & 0.004 & 0.08 & 0.45 & 1.49 & 1.42 & $-$63.5 \\
0119+115 & 0.57 & 1.214 & 0.140 & 0.10 & 0.59 & 1.53 & 1.44 & $-$34.8 \\
0133+476 & 0.859 & 1.882 & 0.014 & 0.11 & 0.56 & 1.55 & 1.41 & $-$61.2 \\
0202+149 & 0.405 & 3.838 & 0.021 & 0.16 & 0.95 & 1.58 & 1.42 & $-$40.5 \\
0202+319 & 1.466 & 0.652 & 0.013 & 0.13 & 0.52 & 1.55 & 1.37 & $-$60.2 \\ 
0212+735 & 2.367 & 2.470 & 0.002 & 0.11 & 0.74 & 1.83 & 1.44 & $-$14.0 \\
0215+015 & 1.715 & 0.430 & 0.090 & 0.08 & 0.50 & 1.74 & 1.43 & $-$26.9 \\
0224+671 & 0.523 & 1.483 & 0.151 & 0.11 & 0.59 & 1.69 & 1.42 & $-$22.1 \\
0234+285 & 1.207 & 2.301 & 0.129 & 0.10 & 1.14 & 1.47 & 1.35 & $-$54.3 \\
0235+164 & 0.94 & 1.510 & 0.031 & 0.06 & 0.45 & 1.51 & 1.38 & $-$35.7 \\
0238$-$084 & 0.0049 & 1.103 & 0.119 & 0.07 & 0.27 & 1.95 & 1.43 & $-$16.5 \\
0300+470 & $-$& 1.174 & 0.065 & 0.08 & 0.35 & 1.47 & 1.41 & $-$36.4 \\
0333+321 & 1.263 & 3.014 & 0.074 & 0.11 & 0.75 & 1.50 & 1.36 & $-$41.0 \\
0336$-$019 &0.852 & 2.917 & 0.077 & 0.20 & 0.73 & 1.74 & 1.44 & 2.2 \\
0403$-$132 & 0.571 & 4.188 & 0.150 & 0.34 & 2.00 & 3.62 & 1.44 & 1.3 \\
0420$-$014 & 0.915& 2.898 & 0.079 & 0.11 & 0.72 & 1.72 & 1.41 & $-$4.6 \\
0422+004 & $-$& 1.088 & 0.005 & 0.13 & 1.08 & 1.71 & 1.40 & $-$16.7 \\
0446+112 & $-$ & 1.541 & 0.033 & 0.16 & 0.38 & 1.57 & 1.38 & $-$28.7 \\
0458$-$020 & 2.291 & 1.029 & 0.174 & 0.10 & 0.70 & 1.78 & 1.41 & $-$18.9 \\
0528+134 & 2.07 & 2.225 & 0.080 & 0.13 & 0.66 & 1.62 & 1.36 & $-$33.2 \\
0529+075 & 1.254 & 1.532 & 0.138 & 0.09 & 0.76 & 1.70 & 1.35 & $-$26.6 \\
0529+483 & 1.162 & 0.647 & 0.026 & 0.10 & 0.39 & 1.50 & 1.36 & $-$23.9 \\
0552+398 & 2.363 & 1.545 & 0.003 & 0.15 & 0.61 & 1.50 & 1.34 & $-$10.9 \\
0605$-$085 & 0.872 & 1.184 & 0.140 & 0.16 & 0.70 & 1.99 & 1.37 & $-$2.8 \\
0607$-$157 & 0.324 & 3.023 & 0.001 & 0.18 & 0.60 & 2.27 & 1.36 & $-$1.2 \\
0642+449 & 3.408 & 0.648 & $<$0.0002 & 0.08 & 0.39 & 1.50 & 1.34 & $-$13.7 \\
0648$-$165 & $-$ & 2.106 & 0.031 & 0.16 & 0.63 & 2.34 & 1.34 & $-$8.3 \\
0716+714 & $-$ & 0.662 & 0.399 & 0.07 & 0.39 & 2.00 & 1.46 & 61.7 \\
0836+710 & 2.218 & 3.238 & 0.175 & 0.19 & 12.65 & 2.24 & 1.43 & 80.9 \\
1222+216 & 0.435 & 0.992 & 1.061 & 0.08 & 0.47 & 1.45 & 1.41 & 24.4 \\
1308+326 & 0.997 & 1.331 & 0.066 & 0.19 & 0.65 & 1.43 & 1.37 & $-$46.2 \\
1324+224 & 1.4 & 1.134 & 0.005 & 0.15 & 0.90 & 1.44 & 1.41 & $-$15.0 \\
1417+385 & 1.832 & 0.514 & 0.008 & 0.07 & 0.40 & 1.46 & 1.37 & $-$42.4 \\
1502+106 & 1.833 & 1.811 & 0.044 & 0.08 & 0.45 & 1.62 & 1.39 & $-$30.3 \\
1538+149 & 0.605 & 1.559 & 0.178 & 0.09 & 0.36 & 1.65 & 1.37 & $-$41.1 \\
1803+784 & 0.68 & 1.979 & 0.025 & 0.11 & 0.79 & 2.93 & 1.47 & 50.7 \\
1849+670 & 0.657 & 0.470 & 0.101 & 0.12 & 0.70 & 3.39 & 1.50 & 55.9 \\
1928+738 & 0.303 & 3.214 & 0.354 & 0.16 & 1.28 & 3.50 & 1.40 & 43.7 \\
2145+067 & 0.999 & 2.859 & 0.034 & 0.16 & 0.71 & 1.73 & 1.45 & 35.4 \\
2155$-$152 & 0.672 & 2.654 & 0.355 & 0.16 & 0.65 & 2.29 & 1.48 & 18.9 \\
2200+420 & 0.0686 & 1.988 & 0.018 & 0.07 & 0.39 & 1.52 & 1.45 & 87.9 \\
2201+171 & 1.076 & 0.843 & 0.106 & 0.06 & 0.41 & 1.63 & 1.44 & 49.9 \\
2201+315 & 0.298 & 1.531 & 0.327 & 0.11 & 0.61 & 1.57 & 1.43 & 74.0 \\
2209+236 & 1.125 & 0.428 & $<$0.0002 & 0.07 & 0.32 & 1.57 & 1.43 & 58.7 \\
2216$-$038 & 0.901 & 1.745 & 0.325 & 0.15 & 2.76 & 1.93 & 1.47 & 27.3 \\ 
2223$-$052 & 1.404 & 6.935 & 0.271 & 0.32 & 2.69 & 1.94 & 1.45 & 26.3 \\
2227$-$088 & 1.562 & 0.924 & 0.011 & 0.07 & 0.37 & 2.08 & 1.46 & 24.7 \\
2230+114 & 1.037 & 6.789 & 0.341 &  0.38 & 2.63 & 1.70 & 1.46 & 39.8 \\
2243$-$123 & 0.63 & 2.265 & 0.032 & 0.10 & 0.56 & 2.19 & 1.46 & 20.1 \\
2251+158 & 0.859 & 14.035 & 0.878 & 0.44 & 4.14 & 1.65 & 1.44 & 37.7 \\
2331+073 & 0.401 & 0.605 & 0.042 & 0.06 & 0.30 & 1.69 & 1.45 & 36.2 \\
2345$-$167 & 0.576 & 1.951 & 0.183 & 0.08 & 0.57 & 2.27 & 1.45 & 17.8 \\
2351+456 & 1.986 & 2.304 & 0.049 & 0.05 & 0.56 & 1.54 & 1.39 & $-$65.5 \\
\enddata 
\tablecomments{Columns are as follows: (1) IAU Name (B1950.0); (2) Optical redshift of host galaxy from NED ;(3) Fitted Gaussian core flux density in Jy; (4) Extended flux density of the source in Jy; (5) R.M.S. noise of the image in mJy $\rm{beam}^{-1}$; (6) Minimum  contour level of the image in mJy $\rm{beam}^{-1}$; (7) FWHM major axis of the restoring beam in arcseconds; (8) FWHM minor axis of the restoring beam in arcseconds; (9) Major axis position angle of the restoring beam in degrees.}

\end{deluxetable} 

\clearpage
\figcaption{1.4 GHz VLA$-$A configuration images of AGN in the 
MOJAVE sample. Each panel contains a Stokes I parameter image of the
source, with contours in successive integer powers of two times the lowest 
contour (see Table 2).  A single negative contour equal in magnitude to
the lowest contour is indicated with dashed lines.}

\figcaption{Distribution of 1.4 GHz extended flux density for 57 AGN in the MOJAVE sample.}

\figcaption{Distribution of core$-$to$-$extended luminosity ratio (source frame) for 57 
AGN in the MOJAVE sample.}

\clearpage

\begin{figure}
\plotone{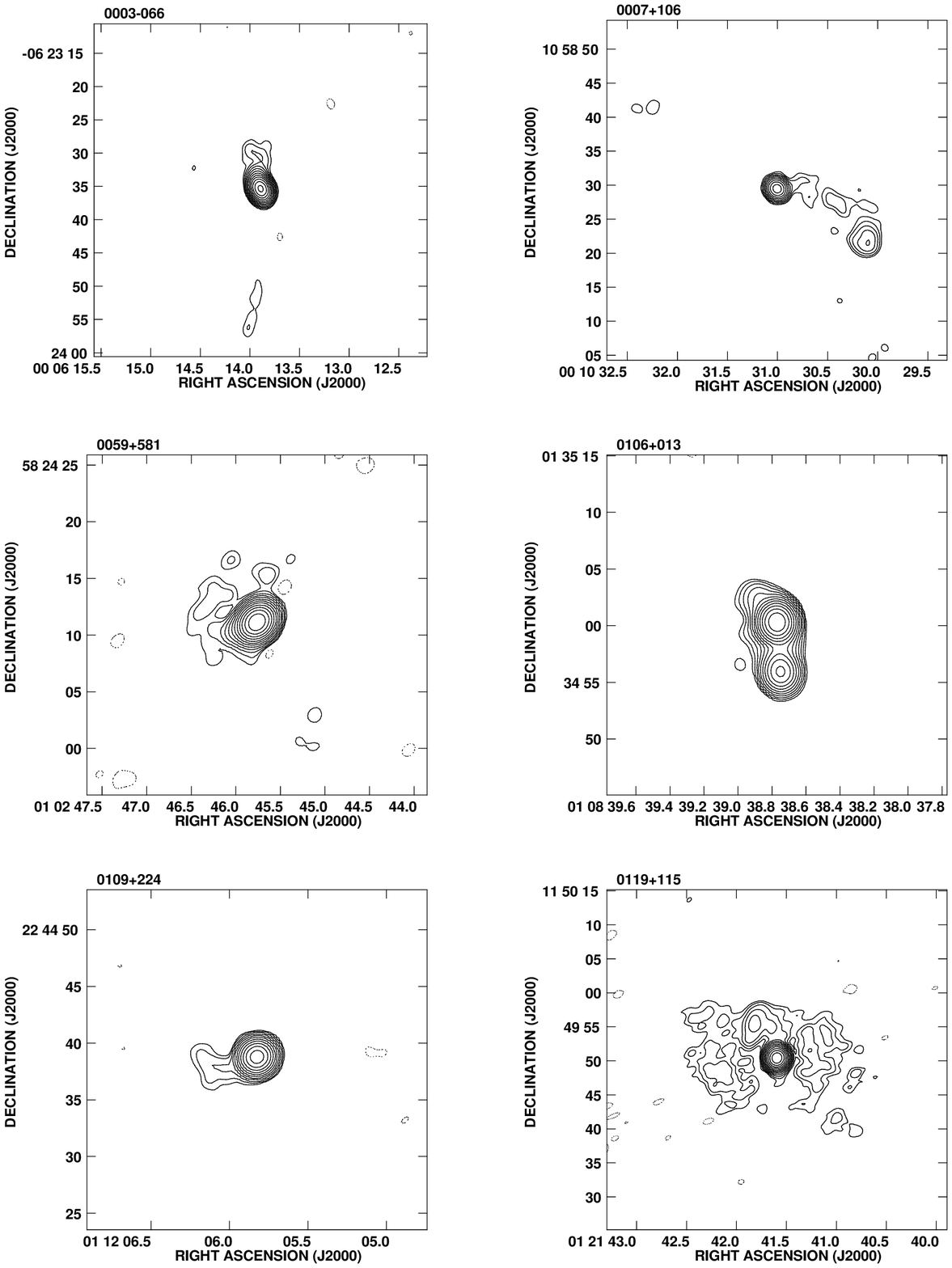}
\centerline{f1a.eps}
\end{figure}
\clearpage

\begin{figure}
\plotone{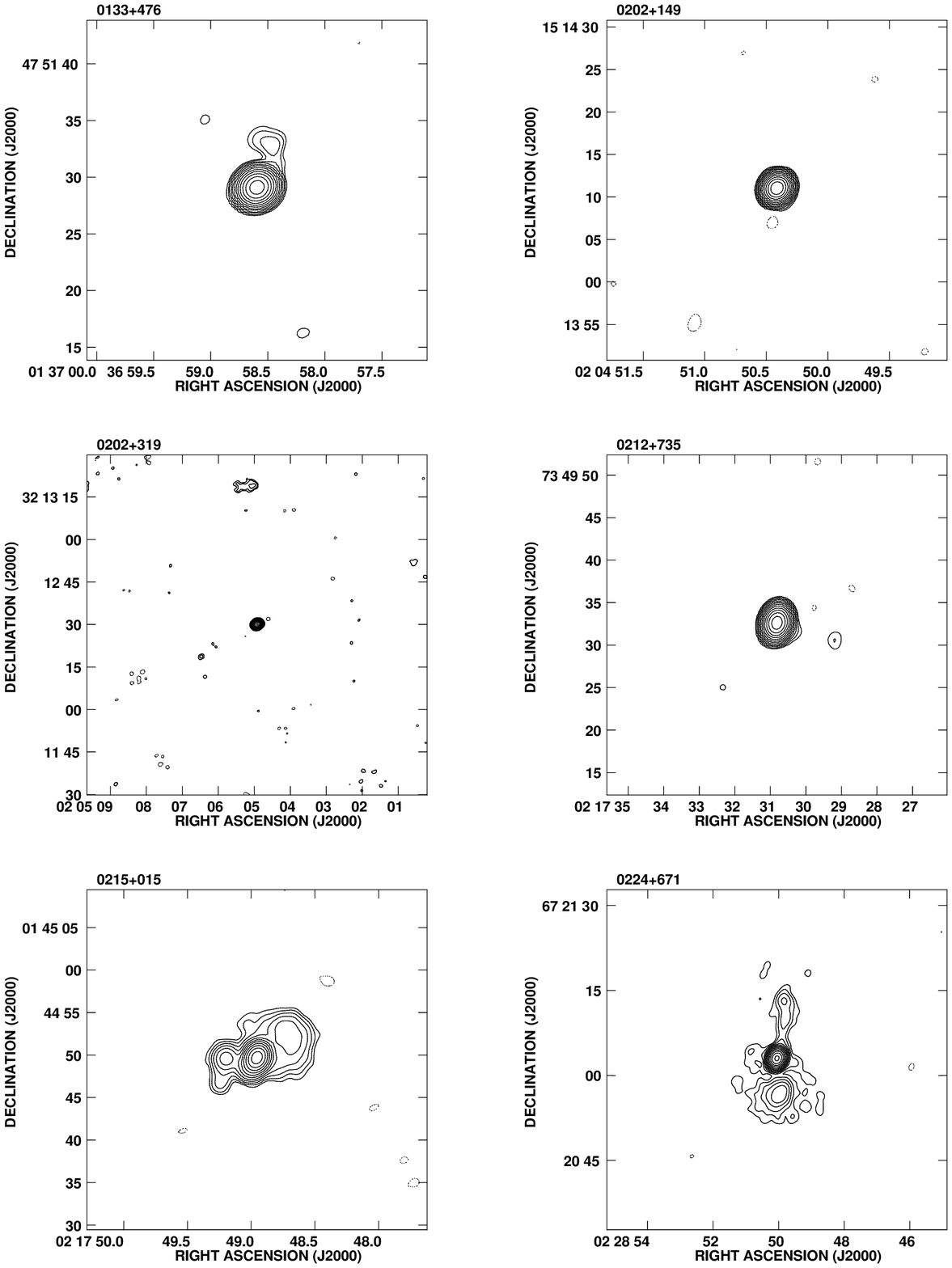}
\centerline{f1b.eps}
\end{figure}
\clearpage

\begin{figure}
\plotone{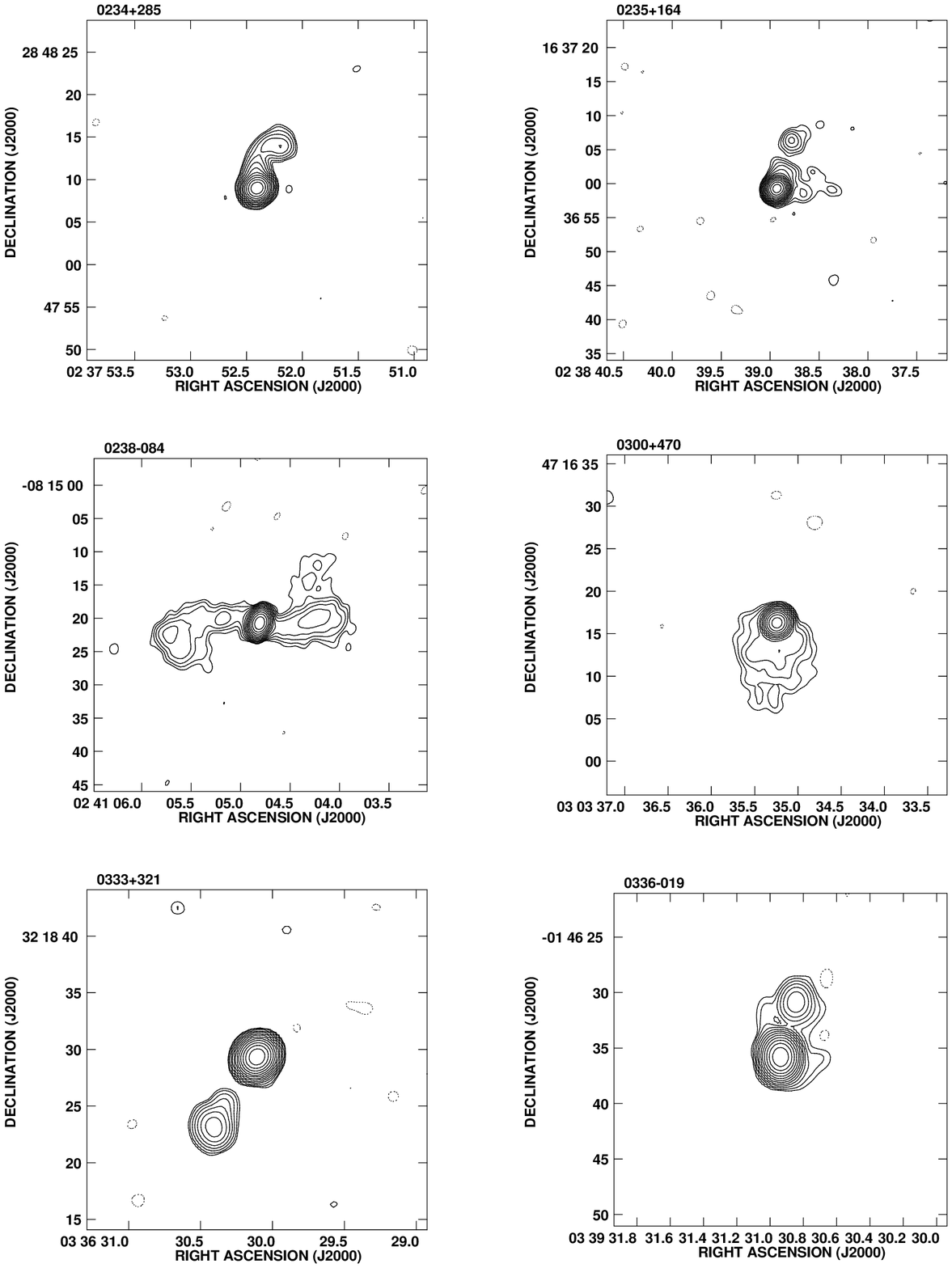}
\centerline{f1c.eps}
\end{figure}
\clearpage

\begin{figure}
\plotone{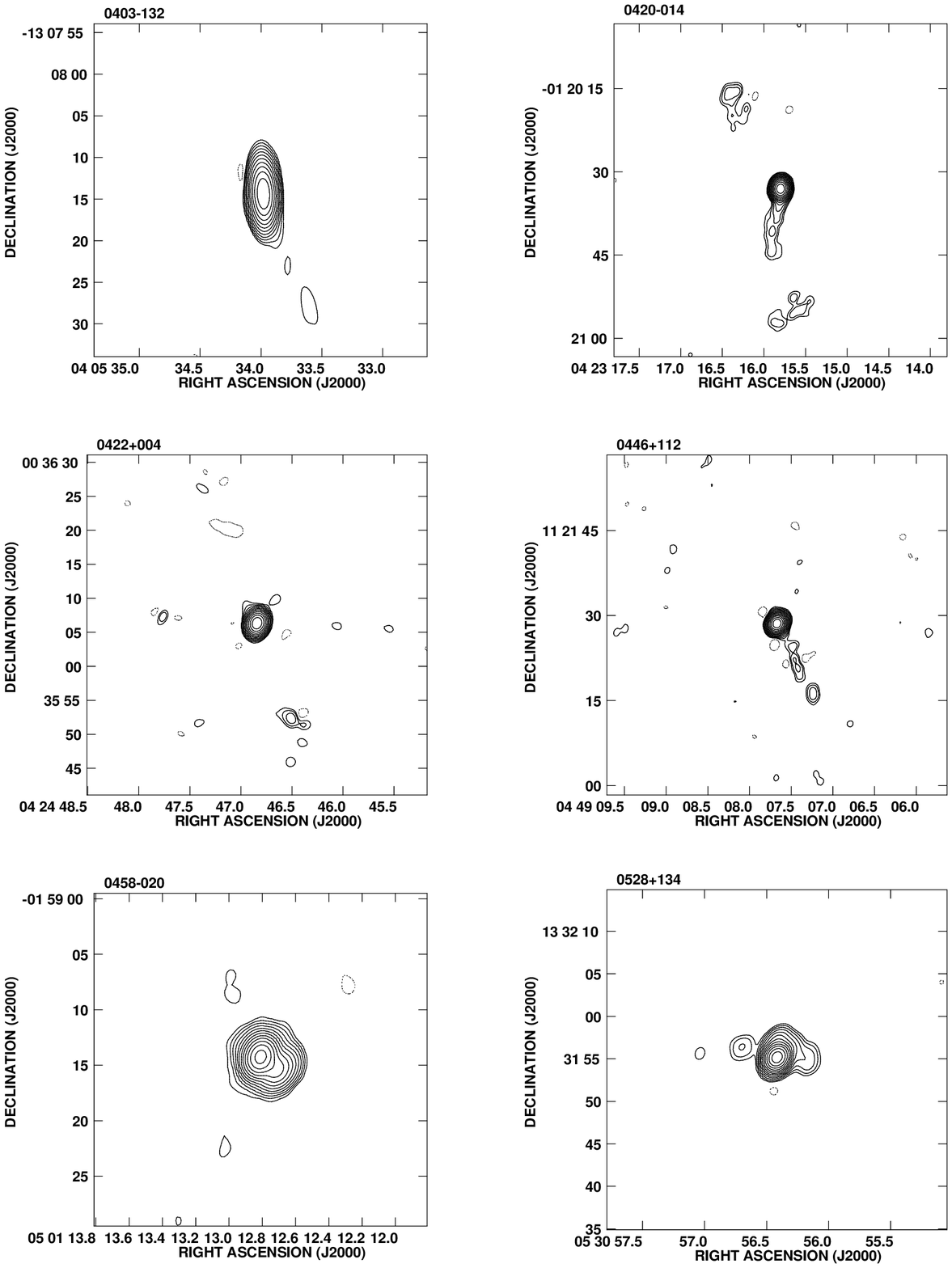}
\centerline{f1d.eps}
\end{figure}
\clearpage

\begin{figure}
\plotone{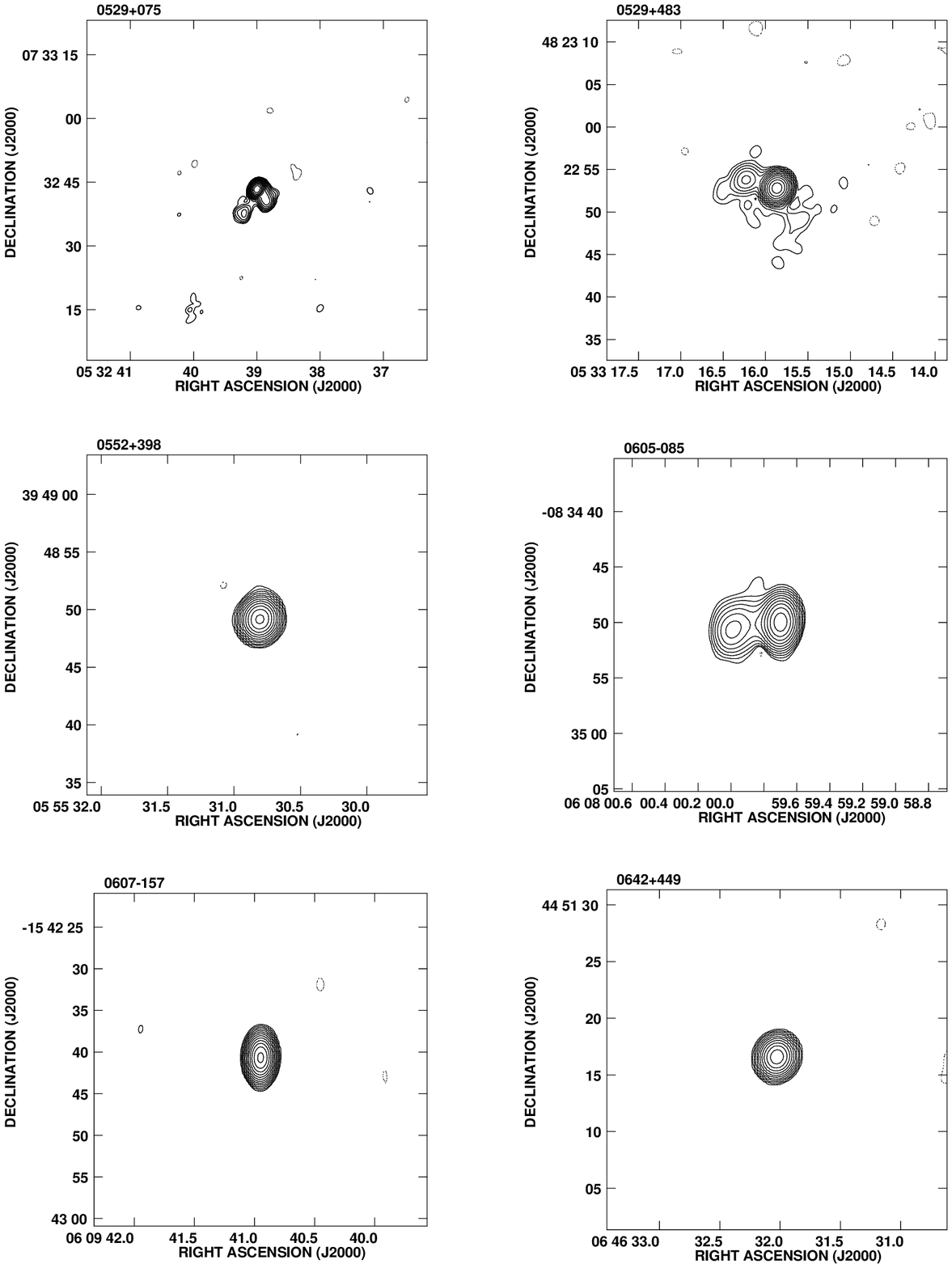}
\centerline{f1e.eps}
\end{figure}
\clearpage

\begin{figure}
\plotone{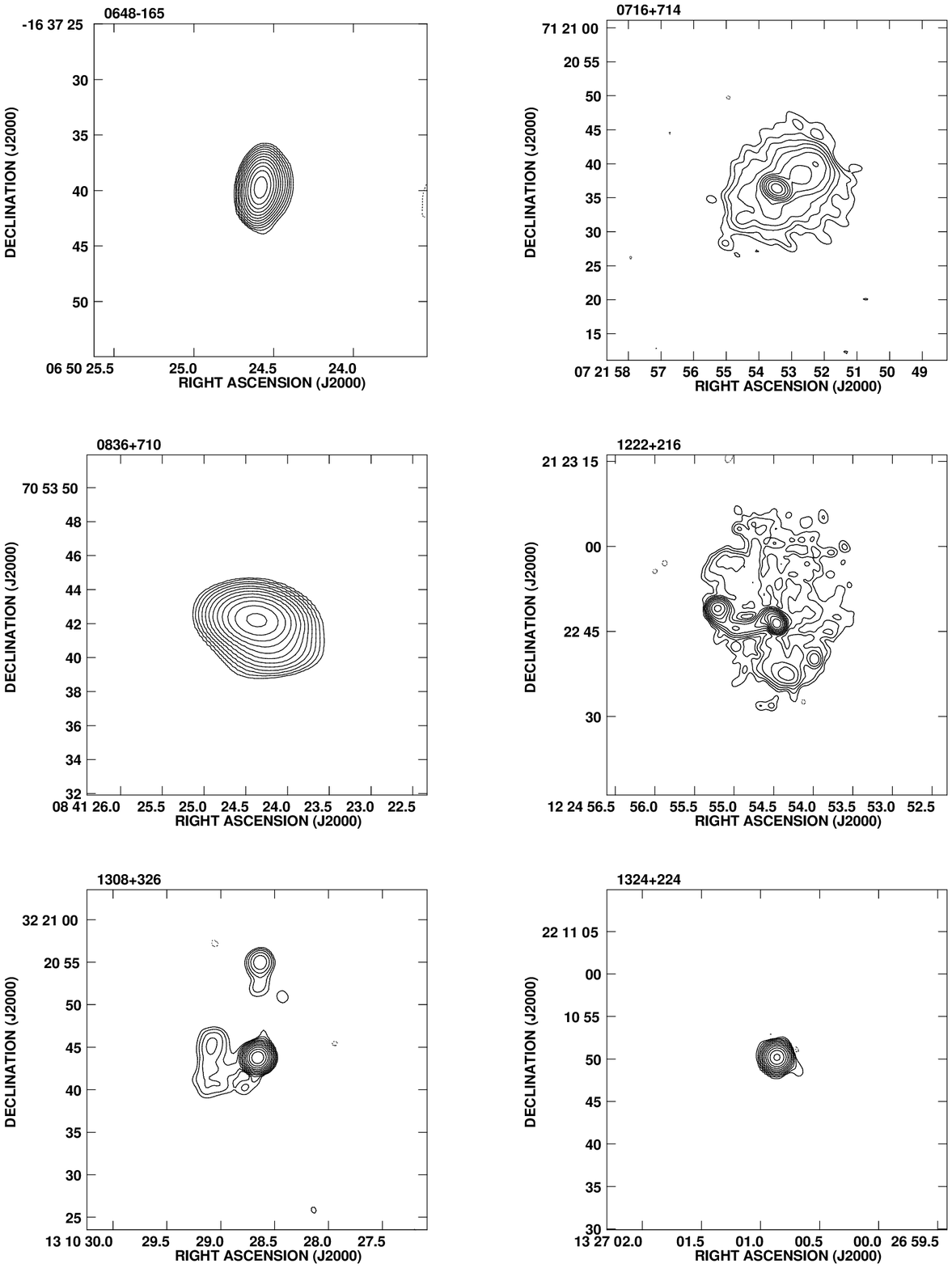}
\centerline{f1f.eps}
\end{figure}
\clearpage

\begin{figure}
\plotone{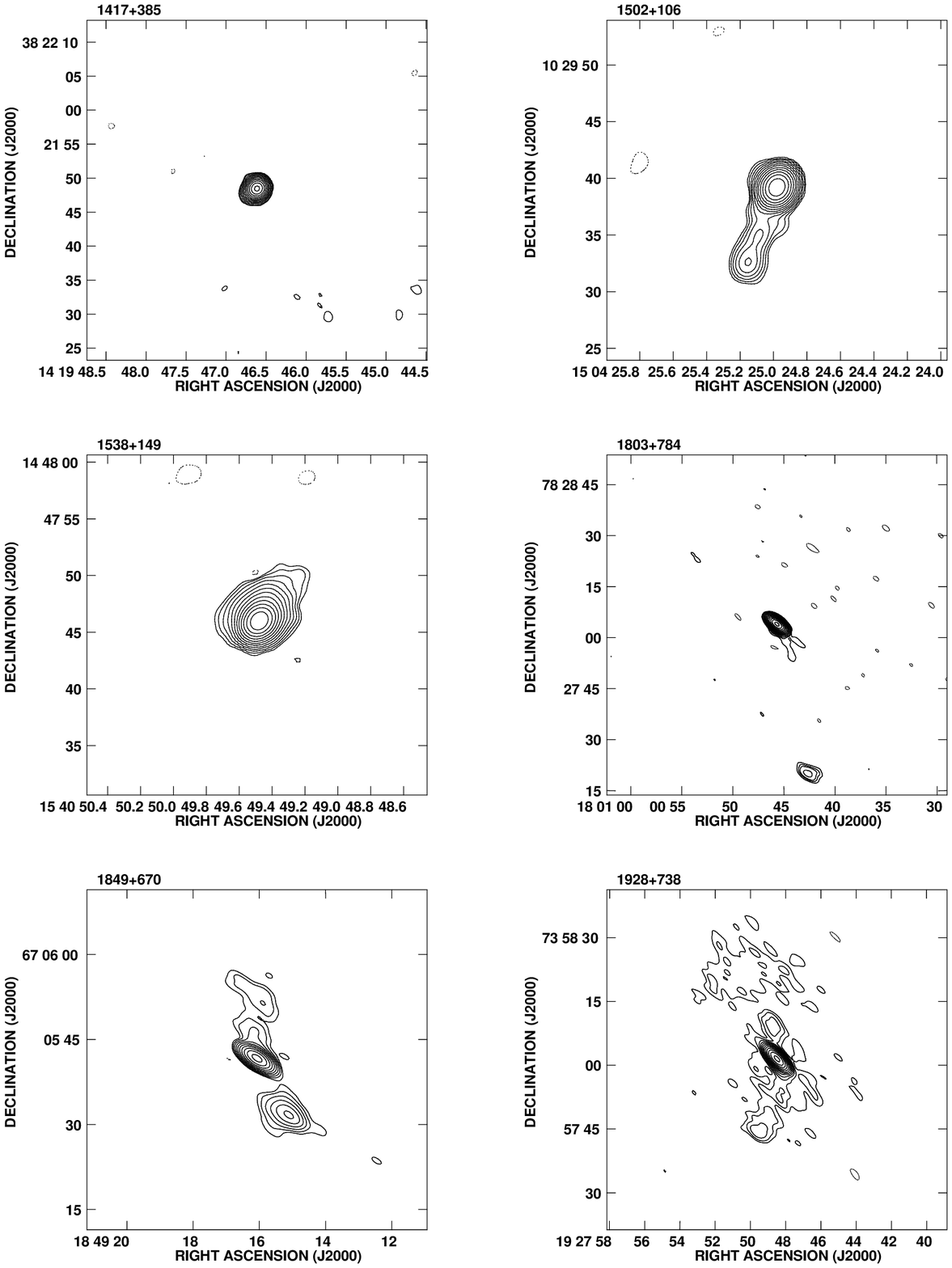}
\centerline{f1g.eps}
\end{figure}
\clearpage

\begin{figure}
\plotone{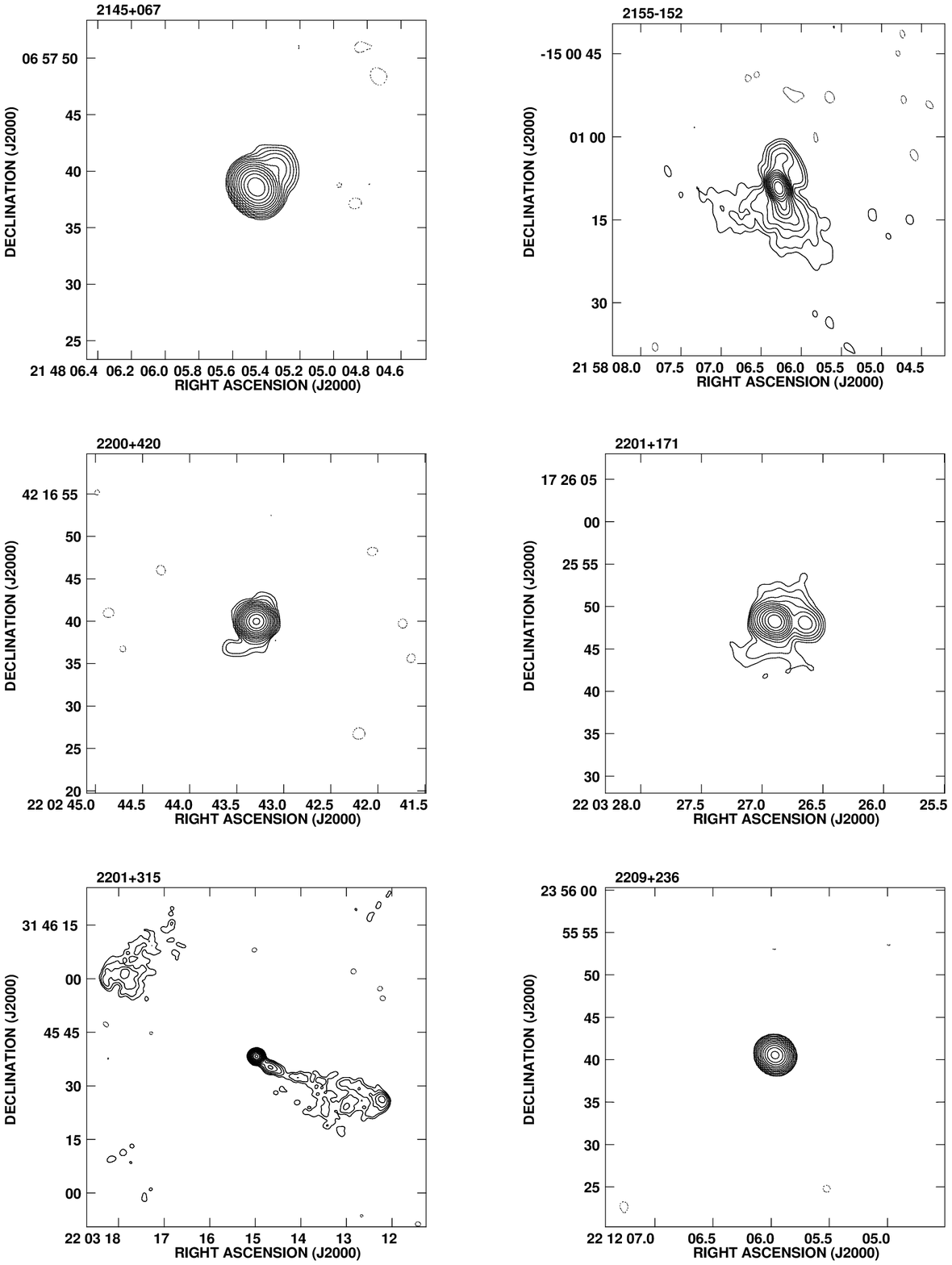}
\centerline{f1h.eps}
\end{figure}
\clearpage

\begin{figure}
\plotone{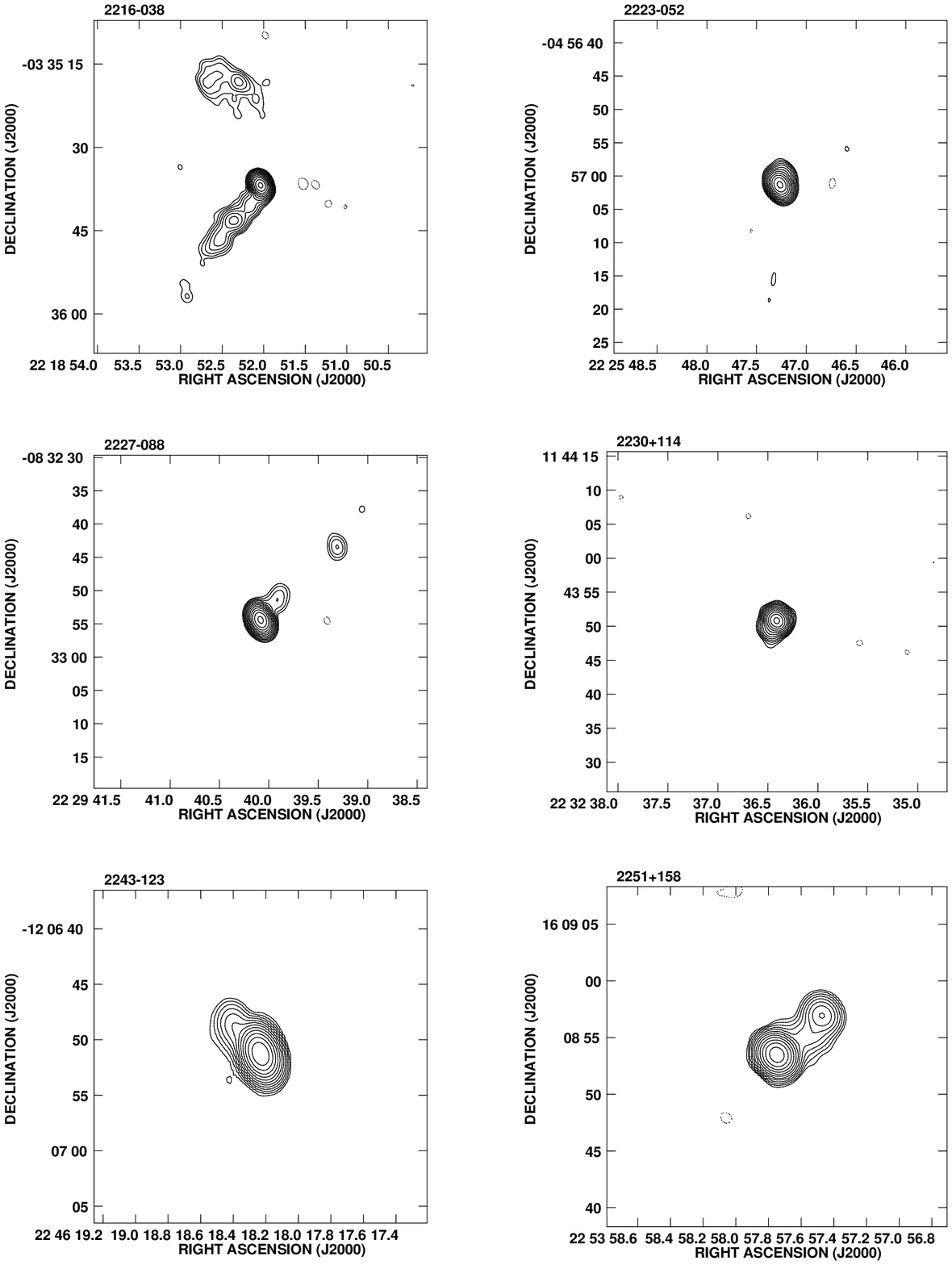}
\centerline{f1i.eps}
\end{figure}
\clearpage

\begin{figure}
\plotone{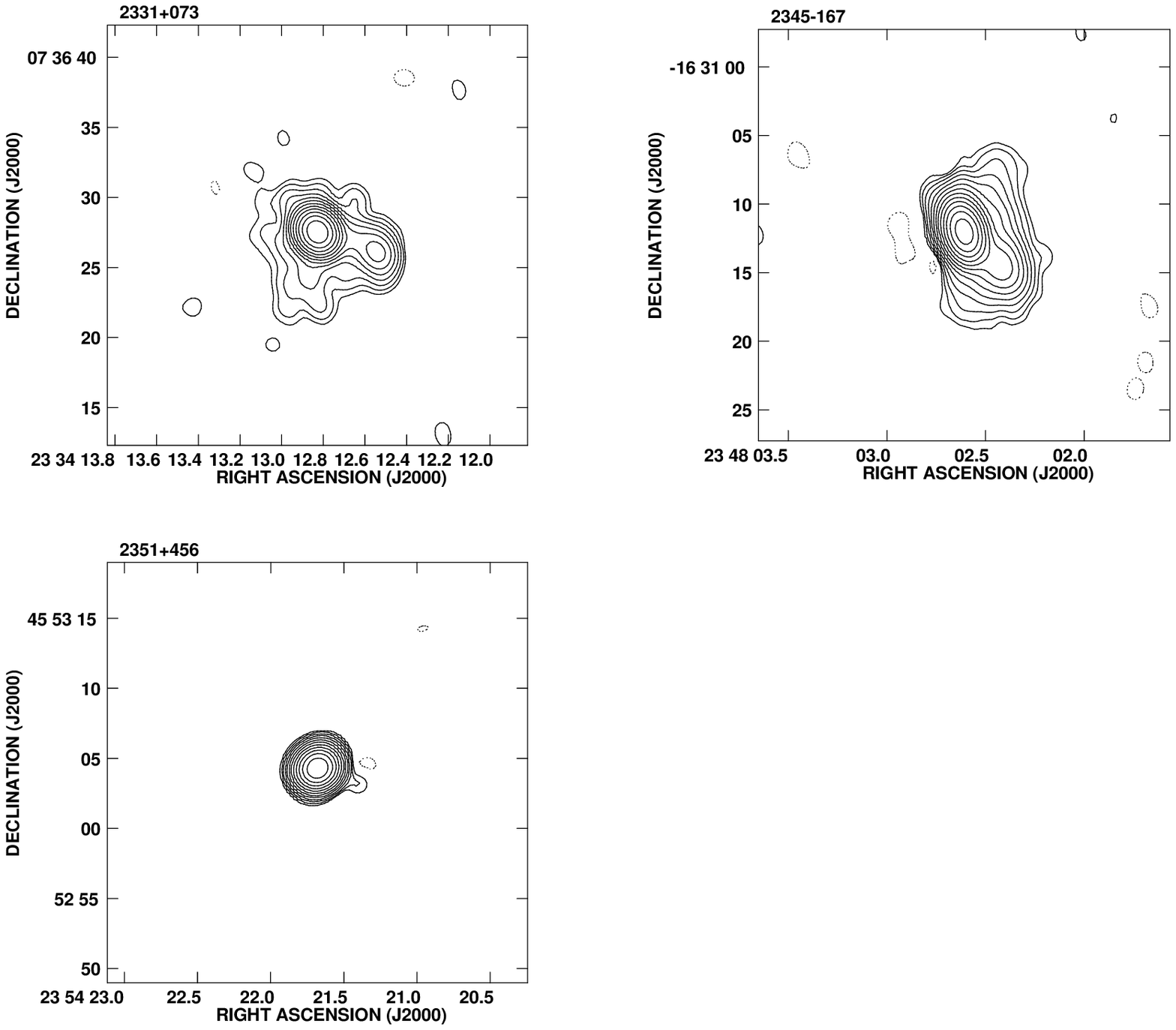}
\centerline{f1j.eps}
\end{figure}
\clearpage

\begin{figure}
\plotone{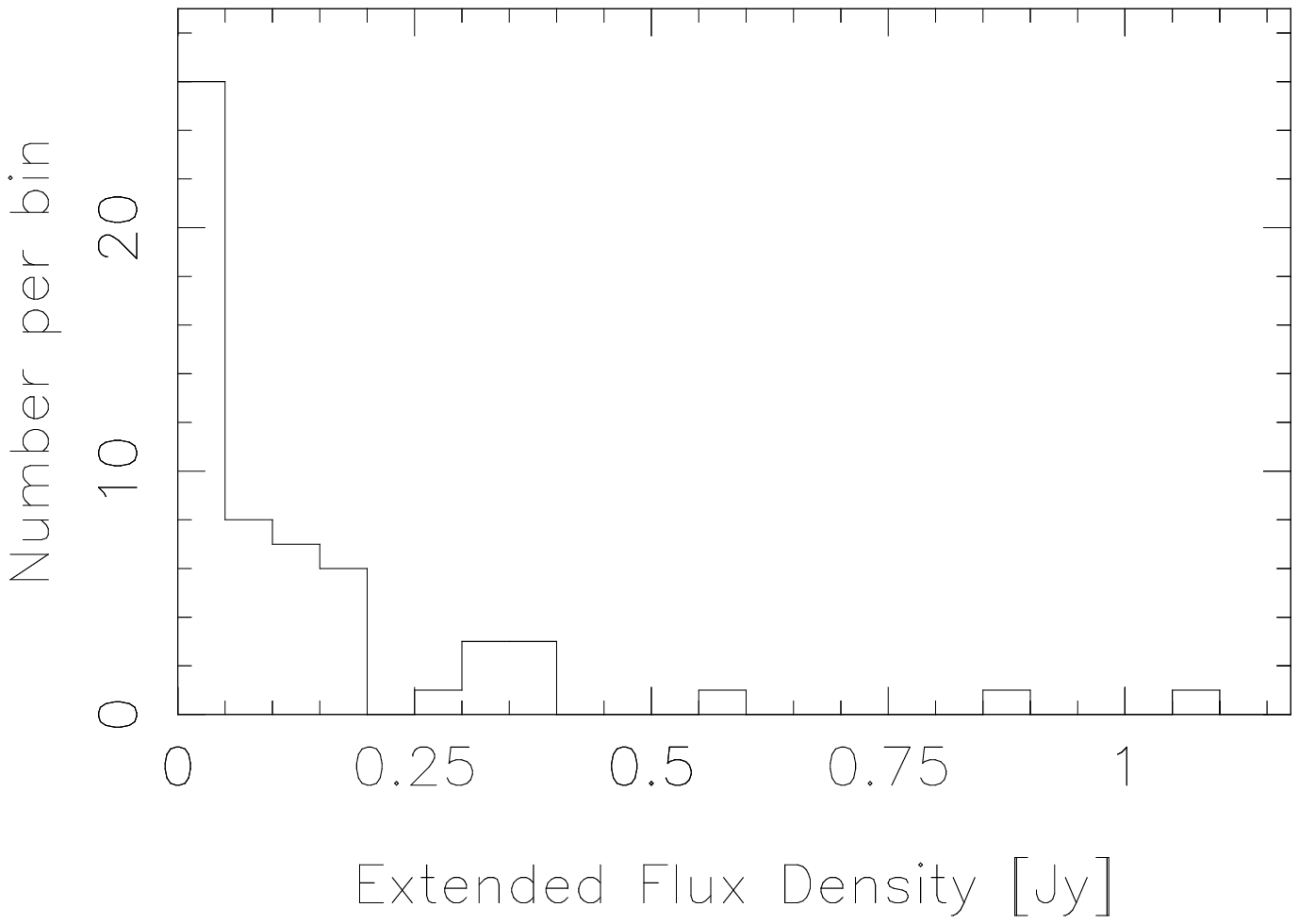}
\centerline{f2.eps}
\end{figure}
\clearpage

\begin{figure}
\plotone{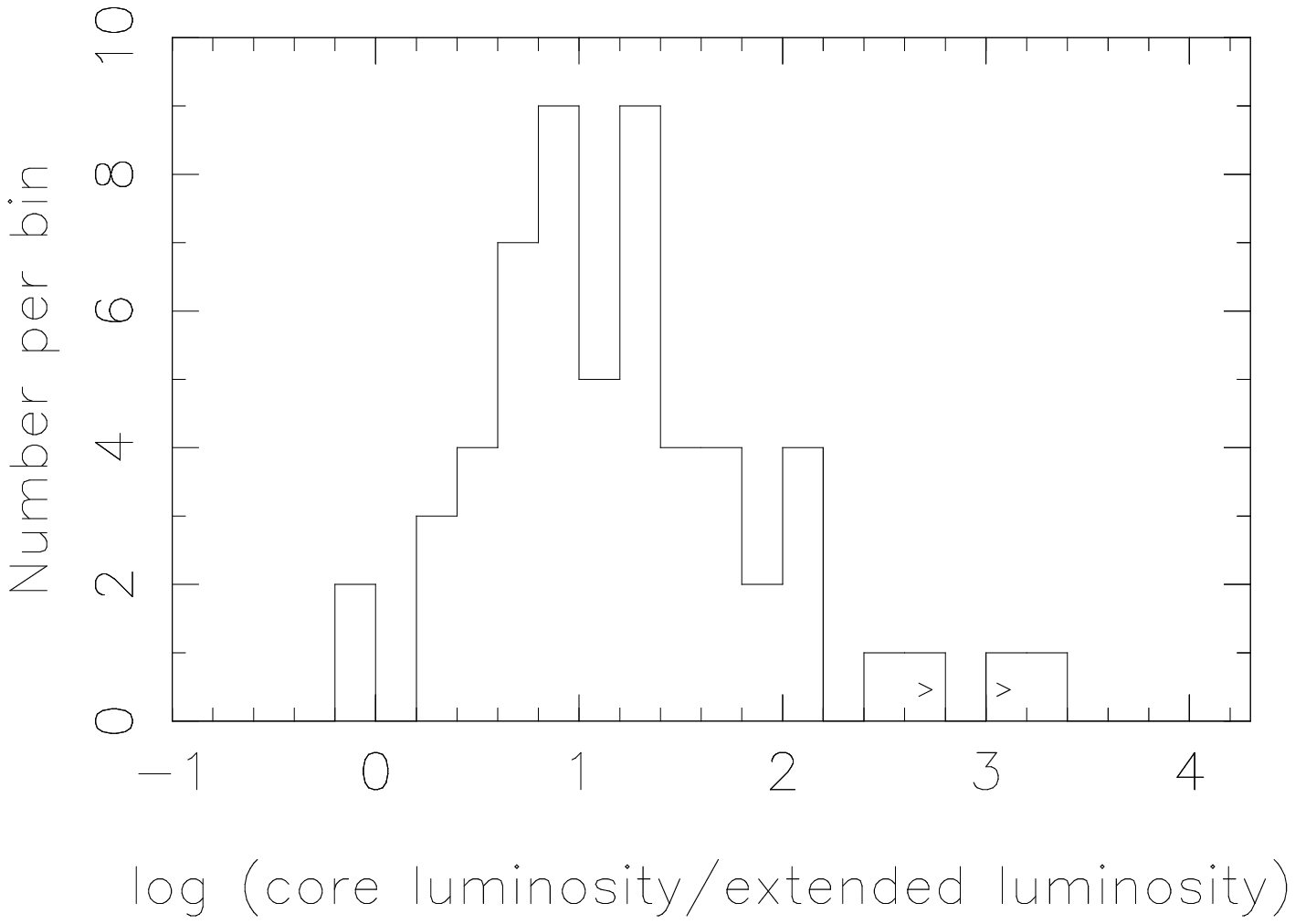}
\centerline{f3.eps}
\end{figure}

\end{document}